\documentclass[twocolumn]{aastex63}
\usepackage{graphicx}
\usepackage{amssymb}
\usepackage{amsmath}
\usepackage{hyperref}


\def\gtrsim{\mathrel{\hbox{\rlap{\hbox{\lower4pt\hbox{$\sim$}}}\hbox{$>$}}}}
\def\lesssim{\mathrel{\hbox{\rlap{\hbox{\lower4pt\hbox{$\sim$}}}\hbox{$<$}}}}

\begin{document}

\title{A lack of 9-s periodicity in the follow-up NuSTAR observation of LS 5039}

\author{Oleg Kargaltsev}
\affiliation{Department of Physics, The George Washington University, 725 21st St. NW, Washington, DC 20052}
\author{Jeremy Hare}
\affil{Astrophysics Science Division, NASA Goddard Space Flight Center, 8800 Greenbelt Rd, Greenbelt, MD 20771, USA}
\affiliation{Center for Research and Exploration in Space Science and Technology, NASA/GSFC, Greenbelt, Maryland 20771, USA}
\affiliation{The Catholic University of America, 620 Michigan Ave., N.E. Washington, DC 20064, USA}
\author{Igor Volkov}
\affiliation{Department of Physics, The George Washington University, 725 21st St. NW, Washington, DC 20052}
\author{Alexander Lange}
\affiliation{Department of Physics, The George Washington University, 725 21st St. NW, Washington, DC 20052}

\email{kargaltsev@gwu.edu}

\begin{abstract}

The Nuclear Spectroscopic Array (NuSTAR) observed the gamma-ray binary LS 5039 for a second time in order to check for the presence of a periodic signal candidate found in the data from the previous NuSTAR observation. We do not detect the candidate signal in the vicinity of its previously reported frequency, assuming the same orbital ephemeris as in our previous paper. This implies that   the previously reported periodic signal candidate was  a noise fluctuation. We also perform a comparison of the lightcurves from the two NuSTAR observations and the joint spectral fitting. Our spectral analysis confirms the phase-dependence found from a single NuSTAR observation at a higher significance level.

\end{abstract}

\section{Introduction}

High-mass gamma-ray binaries (HMGBs) are systems consisting of a high-mass star being orbited by a compact object. The compact object is typically thought to be either a neutron star (NS) or black hole (BH). 
Two Galactic HMGBs (namely, PSR B1259-63 and PSR J2032+4127)
 host young, rotation-powered pulsars 
 which are expected to  
produce powerful relativistic winds. The interaction between the pulsar and stellar winds accelerates particles   
 responsible for the broadband non-thermal (synchrotron and inverse Compton) emission from these systems.  
 However, 
 in several other known HMGBs 
 the nature of the compact object remains unknown. LS 5039 is one of these systems.

LS 5039 is  composed of a massive ($M\approx  23 M_{\odot}$) O6.5V((f)) type star and
a compact object ($M_{\rm co} > 1.6M_{\odot}$) orbiting the star with a period $P_{\rm orb} = 3.90603 \pm 0.00017$ days (the shortest among all Galactic HGMBs) and orbital inclination angle of $i\sim30^{\circ}$ \citep{2005MNRAS.364..899C,2009ApJ...698...514A,2011MNRAS.411.1293S,2011ASSP...21..559C}. Radio
observations have shown persistent AU-scale asymmetric extended emission around LS 5039, which
have been dubbed ``jets'', leading to its microquasar classification. However, the debate on whether
the compact object is an accreting BH or a pulsar interacting with its surroundings and producing
an extended nebula resembling ``jet'' (on milli-arcsecond scales) is still ongoing (see, e.g.,  \citealt{2013A&ARv..21...64D, 2015CRPhy..16..661D} for reviews).  The interest in this topic was further fueled by the recent report of radio pulsations \citep{2022NatAs...6..698W} from a very similar HMGB LS I +61$^{\circ}$ 303  where  magnetar-like flaring was earlier reported  \citep{2012ApJ...744..106T}. 

  \cite{2011MNRAS.416.1514R} performed a search for pulsations in LS 5039 in the  0.005--175 Hz frequency range  using Chandra X-ray Observatory data  and found none. 
However, recently, an 
intriguing claim has been made by \cite{2020PhRvL.125k1103Y} regarding the compact object in LS 5039.
They reported the discovery of 9 s pulsations in 10-30 keV based on the analysis of NuSTAR and
Suzaku HXD data. The 9-s period is within  the range of spin periods of magnetars (see \citealt{2017ARA&A..55..261K} for a review) and was interpreted as such by \cite{2020PhRvL.125k1103Y}. 
\cite{2021ApJ...915...61V} (hereafter V+21) independently analyzed the same NuSTAR and Suzaku data 
 confirming the 
 excess
Fourier power near 9 s, however, they argued that the significance of this excess is not as high as claimed by \cite{2020PhRvL.125k1103Y}.
Since the discovery of a magnetar in HMGB would be an extraordinary finding, shedding light onto
the mystery of magnetar progenitors and their evolution,
 we obtained 
 another
NuSTAR observation of LS 5039 to confirm or refute the tentative magnetar-like periodicity. Here we report the results of this observation together with the results of NICER observation which was carried out at the same time. In Section \ref{obs_and_dat} we describe the NuSTAR observation and data reduction. In Section \ref{analysis} we describe our analysis and  results. 
We conclude in Section \ref{summary} with a summary of our findings.

\clearpage

\section{Observations and Data Reduction}
\label{obs_and_dat}

\begin{figure*}
\includegraphics[trim={0 0 0 0},width=18.3cm]{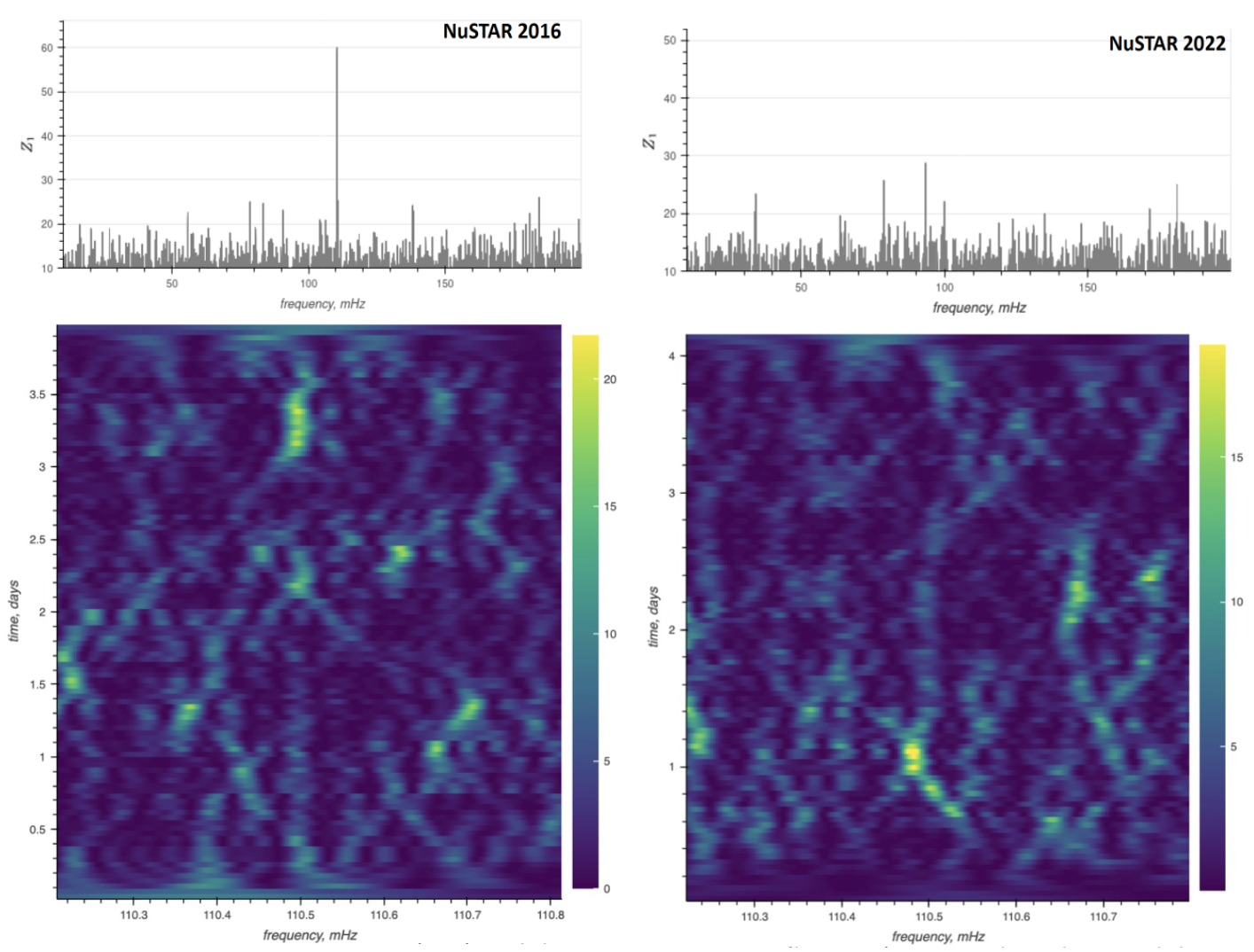}
\caption{ Left and right panels show the total (from the entire observation) Fourier power spectrum (top)  and the dynamic  Fourier power spectrum (bottom; zoomed on the region near the 9-s candidate signal) from the 2016 and 2022 NuSTAR observations, respectively. The detailed description of the left panels in Figure 1, given in V+21, also applies to the right panels (since the length of the observation and the number of photons from the target are very similar for both observations) and, hence, we do not reproduce it here.  
}
\vspace{0.4cm}
\label{obs_orbit}
\end{figure*}

The Nuclear Spectroscopic Array (NuSTAR; \citealt{2013ApJ...770..103H}) observed LS 5039 for $\approx 362$ ks starting on MDJD=59678.94871528 (2022 09 April; ObsID 30701024002). This observation spanned about 1.07 orbits of LS 5039, which has an orbital period of $P_{\rm orb}\approx3.9$ days. The data were processed and reduced using HEASOFT ver. 6.29c and the NuSTAR data analysis software package (NuSTARDAS) version 2.1.1. Prior to performing any analysis, the photon arrival times were corrected to the solar system barycenter using {\tt nuproducts}\footnote{see \url{https://heasarc.gsfc.nasa.gov/lheasoft/ftools/caldb/help/nuproducts.html}}
 and the 
 NuSTAR clock correction file, which improves NuSTARs clock accuracy to about 65 $\mu$s \citep{2021ApJ...908..184B}\footnote{nuCclock20100101v140.fits; see \url{https://nustarsoc.caltech.edu/NuSTAR_Public/NuSTAROperationSite/clockfile.php}}. We further cleaned the data using the flags {\tt saacalc=2}  {\tt saamode=optimized} {\tt tentacle=yes}, which removes time intervals containing enhanced background emission due to the spacecraft's passage through the SAA. After reducing the data about 179 ks of scientific exposure remained. For spectral and timing analysis the source events were extracted from an $r=60''$  circle centered on the source. To produce the background-subtracted spectrum and lightcurve we selected  the  background events  from a $r=90''$ circle placed in a source-free region on the same detector chip.

The details of the 2016 September 1-5 NuSTAR observation and data reduction are given in V+21. The processing was very similar to the one described above.

\section{Data Analysis and Results}
\label{analysis}

Below we describe the orbital variability, 9-s periodic candidate, and orbital phase-resolved spectral analysis of NuSTAR data.

\begin{table}
\caption{Spectral parameters of LS 5039 from joint PL fits to the 2022  NuSTAR and archival Suzaku XIS data} 
\begin{center}
\renewcommand{\tabcolsep}{0.11cm}
\begin{tabular}{lccc}
\tableline  
Phases  & N$_{\rm H}$ & $\Gamma$  & F$_{3-70 \rm \ keV}$ \\
\tableline 
 & 10$^{22}$ cm$^{-2}$ & & 10$^{-11}$ c.g.s.\\
 \tableline 
 0.0-0.1 &   1.1(1) & 1.63(4) & 0.95(5)\\
 0.1-0.2 &   1.21(7) & 1.67(3) & 1.08(4)\\
 0.2-0.3 &   1.24(7) & 1.63(3) & 1.97(7) \\
 0.3-0.4 &   1.26(6) & 1.59(2) & 2.86(7) \\
 0.4-0.5 &   1.14(5) & 1.55(2) & 3.23(8) \\
 0.5-0.6 &   1.3(1) & 1.56(2) & 3.76(8) \\
 0.6-0.7 &   1.24(6) & 1.56(2) & 3.74(8) \\
 0.7-0.8 &   1.18(7) & 1.58(2) & 3.19(9) \\
 0.8-0.9 &   1.17(9) & 1.58(3) & 2.09(7)\\
 0.9-1.0 &   1.07(7) & 1.58(3) & 1.51(6) \\
\tableline 
\end{tabular} 
\end{center}
\tablecomments{ 
The $1\sigma$ uncertainties are shown in parentheses  (see also Figure \ref{fit_orbit}). The XSPEC's  {\tt tbabs} interstellar absorption model  was used.}
\label{tab:1}
\end{table}

\subsection{The 9-s period candidate}

We search for the  previously reported $P=9.05$ s periodic signal  candidate using the same approach as the one described in detail in  V+21. The top panels of Figure 1 show the distribution of the  total (from the entire observation) Rayleigh ($Z_1^2$ statistic; \citealt{1983A&A...128..245B})  in the vicinity of the previously reported candidate signal, which  stands out  in the left top panel. However, it is not present in the data from the 2022 observation shown in the right top panel. The candidate signal was also seen in the dynamic power spectrum\footnote{We note that the width of the frequency interval  shown in the bottom panels is large enough to capture the drifting signal with a constant $\dot{P}\approx10^{-11}$ s s$^{-1}$ (corresponding to the drift of $\delta\nu\approx0.2$ mHz) which is the highest known value  among known magnetars (see e.g., Figure 1 in \citealt{2022NatAs...6..828C}).}
 (vertical stripe of enhanced brightness)  from 2016 observation (bottom left panel) but is not seen in the data from 2022 (bottom right panel). For both data sets the R\"omer delay correction was applied to the photon arrival times using the binary ephemeris found by V+21 (see Table 1 of V+21).  Therefore, we 
 conclude
 that the  
candidate signal reported from the 2016 data  was  due to a noise fluctuation. 
Given the large uncertainties of the ephemeris  (see V+21 for discussion), the search for a signal with a strongly drifting (at an unknown rate) frequency is very expensive computationally and unlikely to result in highly significant detection (V+21 attempted such a search in 2016 observation with a negative result). For this reason, we do not carry out it here.   

\subsection{Orbital variability}

\begin{figure}
\includegraphics[trim={0 0 0 0},width=8.5cm]{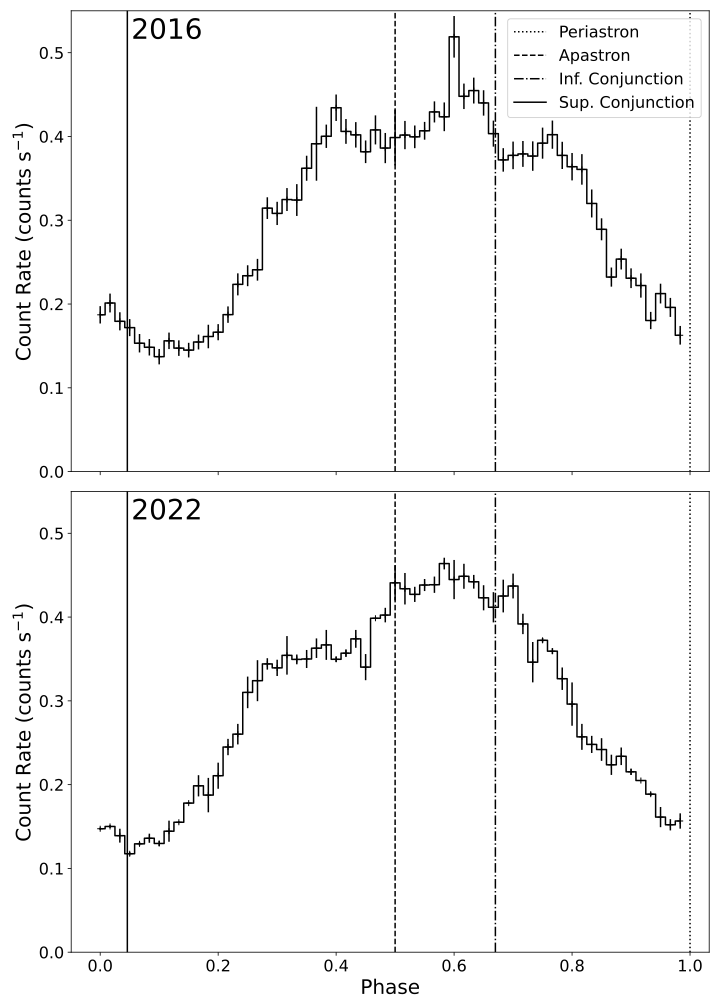}
\caption{ Folded (with a 3.90608-day orbital period) binary lightcurves from the 2016 and 2022 NuSTAR observations in 3-79 keV range. 
}
\label{lightcurves}
\end{figure}

Figure \ref{lightcurves} shows the comparison of lightcurves from two binary cycles corresponding to the 2016 and 2022 observations.  
Photon arrival times were processed with {\tt nuproducts} which uses the \emph{nulccorr} task to apply the ``livetime'' correction and creates the binned lightcurves (we selected the binsize of 5.6 ks which results in  $\approx$60 bins for a single observation lightcurve). We then calculated binary phases corresponding to each lightcurve bin using the 3.90608-day orbital period with a reference (zero phase) epoch of MJD 52825.48  \citep{2009ApJ...698...514A}.

Although the overall shape appears to be preserved, there are some changes at the  upper part of the lightcurve. In 2016 observation the top was flat with some short and spiky episodes of emission while in 2022 it appears to have evolved into a more asymmetric  (``shoulder-head'') shape and became smoother. The observed changes are reminiscent of those shown in Figure 7 of \cite{2023ApJ...948...77Y} comparing the 2007 Suzaku XIS  and 2016 NuSTAR lightcurves in 3-10 keV. We, however, note that the latter comparison involved detectors with different dependencies of their effective areas on energy and, hence, the difference may partly be due to the phase-varying spectrum (see Figure \ref{fit_orbit}).  Also, the lightcurve minimum appears somewhat narrower in the 2022 observation. According to \cite{2009ApJ...698...514A,2011ASSP...21..559C} the uncertainty of LS 5039 orbital period measurement is $\approx 0.0001$ days (at 1$\sigma$ level) which translates into a maximum shift of $\approx0.05$ in the binary phase over the 5.6-yr period between the two NuSTAR observations (i.e. 2-3 bins for the lightcurves shown in Figure \ref{lightcurves}).

\subsection{Spectrum}

\begin{figure}
\includegraphics[trim={0 0 0 0},width=9.0cm]{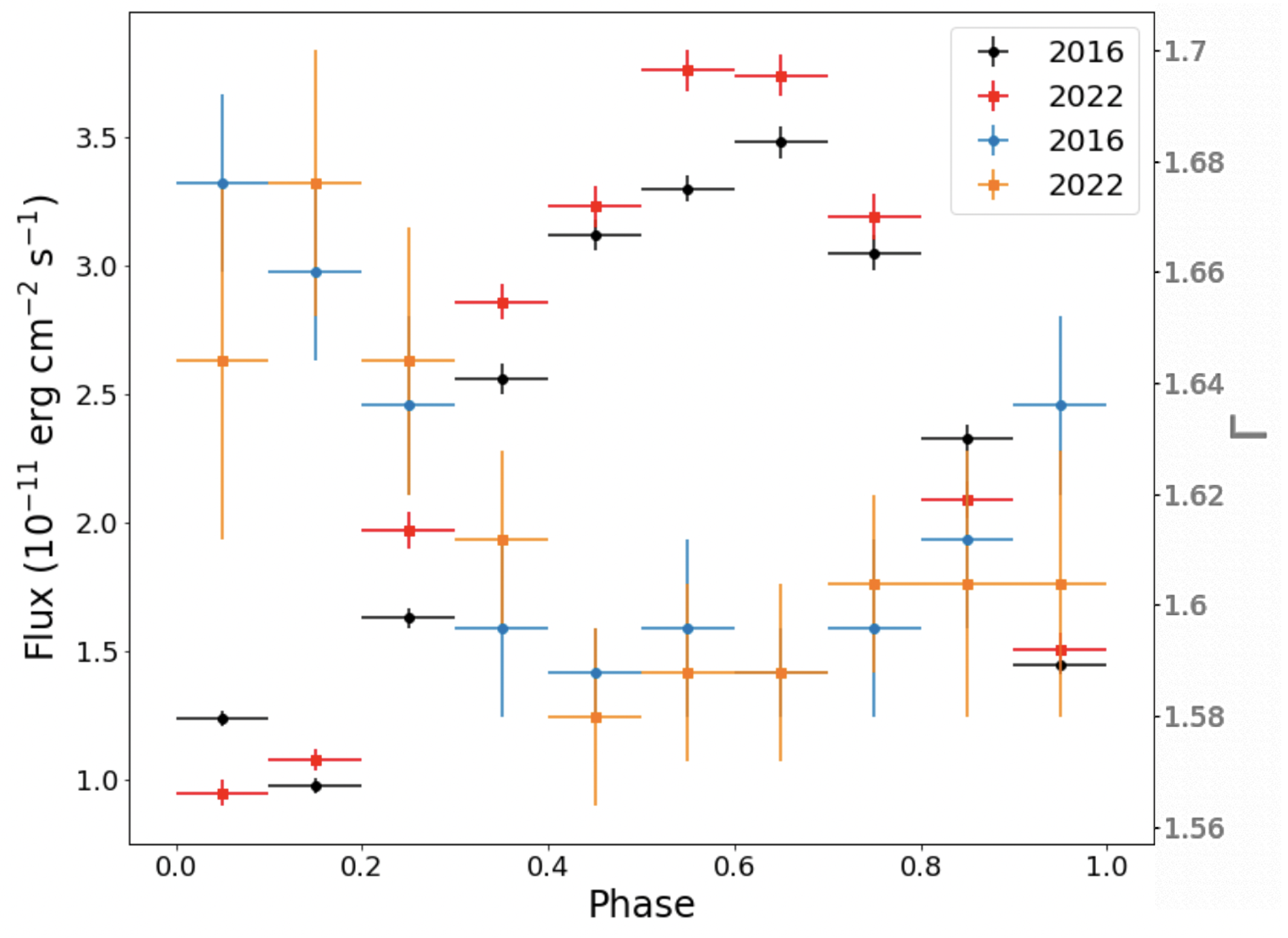}
\caption{ Plot of flux
 and photon index ($\Gamma$) as a function of phase for the 2016 and 2022 NuSTAR observations. 
NuSTAR is jointly fitted with Suzaku XIS data in both cases. 
 Black and red errorbars are the flux values (in 3-70 keV) for 2016 and 2022  NuSTAR data, respectively. 
Blue and orange errorbars correspond to the photon index (shown on the right vertical axis) obtained using  2016 and 2022  NuSTAR data, respectively. The data points that use 2016 NuSTAR data are adopted from V+21.} 
\label{fit_orbit}
\end{figure}

Following V+21 we divided the orbital period into 10  bins and fitted jointly 2022 NuSTAR (3-70 keV) and Suzaku XIS (0.7–10 keV) data. The Suzaku XIS data, obtained in 2007, are the same as those used for the joint fitting in V+21 (see V+21 for more detailed description). The NuSTAR spectral extraction was performed the same way as in V+21 for the 2016 data.  The results of the fit with absorbed\footnote{We used {\tt tbabs} Galactic absorption model   with  {\tt wilms} abundances \citep{2000ApJ...542..914W} and performed the fits in XSPEC v.12.12.0 \citep{1996ASPC..101...17A}. } power-law (PL) model are shown in Table \ref{tab:1} and Figure  \ref{fit_orbit}. The PL fit quality was good for all phase bins. The new data confirm the previously found anti-correlation between the flux and photon index and show that this picture remains stable on the timescale of years.  Our results are also in general agreement with the earlier published analysis of \cite{2005ApJ...628..388B,2009ApJ...697L...1K,2009ApJ...697..592T} who used  data from ASCA, BeppoSAX, RXTE, XMM-Newton, CXO, and Suzaku. 

We  also used phase-integrated spectra from both NuSTAR observations to look for any evidence of spectral cutoff but, similar to V+21, did not find one. Since the variations in the photon index with the phase are rather small, the PL model fits well the phase-integrated spectrum ($\chi^{2}_nu=0.95$ for 1939 d.o.f.) and the fit shows no systematic residuals at any energy range including the high-energy end of the spectrum.

\section{Summary}
\label{summary}

We analyzed the second NuSTAR observation of the gamma-ray binary LS 5039 mirroring our analysis of the first NuSTAR observation (presented in V+21). We do not confirm the candidate 9-s period discussed in  \cite{2020PhRvL.125k1103Y} and V+21. Although similar, the orbital lightcurve shows some differences from the previous one which can be explained by the  varying properties of the stellar wind. We re-affirm the previously reported anti-correlation between the flux and PL slope with the latter varying between 1.59 and 1.68.  No evidence of spectral cutoff is seen at the higher-energy end of the NuSTAR band. 

\medskip\noindent{\bf Acknowledgments:}
Support for this work was provided by the National Aeronautics and Space Administration through
the NuSTAR award 80NSSC22K0013. J. H. acknowledges support from NASA under award number 80GSFC21M0002.

\end{document}